\documentclass[runningheads]{llncs}
\usepackage[T1]{fontenc}

\usepackage{amssymb}
\usepackage{amsmath}
\usepackage{graphicx}
\usepackage{hyperref}
\usepackage{color}
\usepackage[table]{xcolor}

\usepackage{algorithm}
\usepackage{algpseudocode}
\usepackage{setspace}
\usepackage{todonotes}
\usepackage{nicematrix}
\usepackage{caption,subcaption}
\usepackage{paralist}

\usepackage{tikz}
\usetikzlibrary{petri,positioning}

\algnewcommand{\LeftComment}[1]{\Statex \(\triangleright\) #1}

\newcommand{\pre}[1]{\bullet{#1}}
\newcommand{\post}[1]{{#1}\bullet}
\newcommand{\larrow}[1]{\ensuremath{\mathrel{\smash{\stackrel{#1}{\longrightarrow}}}}}

\usepackage{dutchcal}

\usepackage{lipsum}
\usepackage{bbding}

\newcommand{\highlight}[2]{\colorbox{#1}{$\displaystyle #2$}}

\begin{document}
\title{Secure Conformance Checking using Token-based Replay and Homomorphic Encryption}
\titlerunning{Secure token-based replay}
\author{
Luis-Armando Rodríguez-Flores\inst{1}\orcidID{0000-0003-0070-1248} \and
Luciano García-Bañuelos\inst{1}\orcidID{0000-0001-9076-903X} \and
Abel Armas-Cervantes\inst{2}\orcidID{0000-0003-0628-2451} \and
Astrid-Monserrat Rivera-Partida\inst{1}\orcidID{0000-0003-0302-9610}
}
\authorrunning{Rodríguez-Flores et al.}
\institute{Tecnologico de Monterrey, Mexico
\email{\{lrodriguez,luciano.garcia,A01324504\}@tec.mx}\\
% \url{http://www.springer.com/gp/computer-science/lncs} 
\and
The University of Melbourne, Australia\\
\email{abel.armas@unimelb.edu.au}
}
\maketitle              % typeset the header of the contribution
%
% \begin{abstract}
% The abstract should briefly summarize the contents of the paper in

% \keywords{Conformance checking, Confidentiality preservation, Homomorphic encryption}
% \end{abstract}

\begin{abstract}
	Conformance checking, one of the main process mining operations, aims to identify discrepancies between a process model and an event log. The model represents the expected behaviour, whereas the event log represents the actual process behaviour as captured in information systems' records. Traditionally, the process model and the event log are both accessible to the business analyst performing the conformance checking. However, in some contexts the log's owner may want to protect critical or sensitive information in the log and still check its conformance with respect to a model belonging to another party. In this paper, we propose a secure approach to conformance checking based on the well-known token-based replay algorithm and homomorphic encryption. An evaluation is performed using a synthetic log, showing the practicality of the proposed technique.
\end{abstract}

\section{Introduction and related work}
\label{sec:introduction}

Conformance checking~\cite{carmona2018conformance} is one of the main operations in process mining. This operation contrasts the observed and expected process behavior to identify undesirable deviations. The expected behavior is given as a process model and the observed behavior is captured as an event log recorded by information systems. \emph{Event logs} are collections of process executions, which are sequences of activity instances (\emph{a.k.a. events}) commonly referred to as \emph{traces}. 
Conformance checking approaches can be classified into three families: rule-based, alignment-based and token-based replay. The work reported here belongs to the last family. 

Token-based replay was introduced in the seminal work~\cite{Rozinat08}. Later, with the appearance of alignment-based techniques~\cite{Adriansyah2014AligningOA}, token-based replay became less used due to its scalability issues. The scalability issues were caused by the original version requiring exhaustive explorations of the reachability graph, resulting in high overhead. Later, Berti et al.~\cite{BertiA19} revisited token-based replay to address the scalability problems and reignite the interest in this approach.
In essence, token-based replay consists of firing traces captured in an event log to determine if they are accurately described by a process model. Specifically in the context of a process described as a Petri net model, a trace is accurately described if, from an initial marking, the trace can be replayed on the Petri net and a desired final marking is reached. If deviations occur, then tokens would remain after the trace has been replayed, or artificial tokens would be artificially added to the Petri net to make it possible for an event in a trace to be fired. %Then, detected deviations are described as tokens -- remaining or introduced -- necessary to replay the event log. 

Traditionally, conformance checking is performed ``in clear'' such that the model and the event log are known to the end-user performing the operation. However, in some scenarios, the process model and event log must be kept private. For instance, in custom manufacturing, manufacturers strive to keep their production details secret, especially those related to the production of custom pieces made for customers developing new high tech products. Still customers would be willing to check the progress on the production of their pieces. The latter calls for a conformance checking technique over encrypted data. %Other scenarios would include benchmarking where companies would not like to disclose their processes while still comparing them against other competitors. 

Some studies have proposed solutions to safeguard confidentiality of event logs using encryption techniques. For instance,~\cite{BurattinCT15} presents a framework that enables outsourcing process mining analyses while hiding sensitive information in event logs through the use of symmetric and homomorphic encryption. Since the structure of the event log is preserved, it is possible to apply most process mining tools to the encrypted datasets. However, solely encrypting event logs may leave them vulnerable to de-anonymization techniques. To mitigate this risk, the work in \cite{RafieiWA19} proposes a confidentiality framework that integrates encryption with abstraction methods (directly follows matrices of activities and resources) to discover process models and social networks from event logs. 

The challenge of ensuring privacy in cross-organizational contexts was introduced in~\cite{LiuDZZLC19}. The approach requires all involved parties sharing public fragments of their process models with a trusted third party, which then merges these fragments into a cross-organizational model. Then, each organization retrieves the fragments relevant to its operations and combines them with its private models. Similarly,~\cite{ElkoumyFDLPW20} proposes a method that allows multiple parties to perform basic process mining operations -- such as process discovery and performance analysis queries -- over encrypted sub-logs belonging to the different parties in such a way that they do not share confidential information with each other. Strategies such as vectorization of event logs and a divide-and-conquer scheme to process sub-logs in parallel were used to address scalability issues.
Despite these advancements, at the time of this work, we are not aware of any other work applying homomorphic encryption in a conformance checking setting.

This paper presents an implementation of the token-based replay approach for conformance checking using homomorphic encryption. \textcolor{black}{In order to do so, it is necessary to re-formulate the approach using matrix multiplication operations. Specifically, this paper presents a new token-based replay formulation using the marking equation.} To show the practicality of this new formulation, it is evaluated using a synthetic event log.

The remainder of the paper is organized as follows. Section~\ref{sec:background} introduces relevant concepts, and definitions used in the rest of the paper. The approach is presented in Section~\ref{sec:approach} and is evaluated in Section~\ref{sec:evaluation}. The evaluation is used a synthetic %\textcolor{red}{and a real-life} 
event log. Finally, Section~\ref{sec:conclusion} presents the conclusions and some directions for future work.

\section{Background}
\label{sec:background}

An important concept used in the rest of the paper is that of multiset, which is a generalization of the notion of set. In a multiset, there are multiple instances of an element. The formal definition is presented next. 

\begin{definition}[Multiset]\label{def:multiset}
Let $X$ be a set. $B \in \mathcal{B}(X) = X \rightarrow \mathbb{N}$ is a multiset over $X$ where each element $x\in X$ appears $B(x)$ times. A relevant operation defined between multisets $B_1,B_2 \in \mathcal{B}(X)$ is that of difference:

\begin{itemize}
    \item (difference) $B' = B_1 \backslash B_2 \Longleftrightarrow B'(x)  = max(B_1(x) - B_2(x), 0) ~\forall x \in X$. We can also say, in the same setting, that $B' = B_1 - B_2$.
\end{itemize}

\end{definition}
    
\subsection{Petri nets}
\label{subsec:nets}

In the remainder of the paper, $\mathcal{A}$ represents the universe of activity names, and $A \subseteq \mathcal{A}$ is a set of activity names. We also consider the symbol $\tau \notin \mathcal{A}$, which represents internal actions that remain hidden to any system observer. With abuse of notation, we define $A^\tau = A \cup \{\tau\}$.

\begin{definition}[(Labeled) Petri Net, Accepting Petri Net]
A \emph{labeled Petri net}, or simply a \emph{net}, is a tuple $N = (P, T, F, \lambda)$, where  $P$ is a finite set of places,  $T$ is a finite set of transitions, with  $P \cap T = \emptyset$, $F \subseteq (P \times T) \cup (T \times P)$ is a set of arcs called flow relations, and $\lambda: T \rightarrow A^\tau$ is a labeling function assigning an activity name to a transition or $\tau$ if the transition is silent. 
A marking $M: P \rightarrow \mathbb{N}_0$ is a function associating a place $p \in P$ with a (natural) number of tokens in $p$. 

%A net system $\mathcal{N} = (N,M_0)$ is a Petri net $N = (P, T, F, \lambda)$ with a fixed initial marking $M_0$.
An \emph{accepting Petri net} $\mathcal{N} = (N,M_0, M_f)$ is a Petri net $N = (P, T, F, \lambda)$ with a fixed initial marking $M_0$ and a final marking $M_f$.

\end{definition}

%\todo{I think the notion of final marking is important for token-based replay right? I introduce the notion of Accepting net as per van de Aaalst}

Places and transitions are conjointly referred to as \emph{nodes}. We write $\pre{y} = \{x \in P\cup T \ | \ (x,y) \in F\}$ and $\post{y} = \{z \in P\cup T \ | \ (y,z) \in F\}$ to denote the \emph{preset} and \emph{postset} of node $y$, respectively. 
%$F^+$ and $F^*$ denote the irreflexive and reflexive transitive closure of the flow relation, respectively.

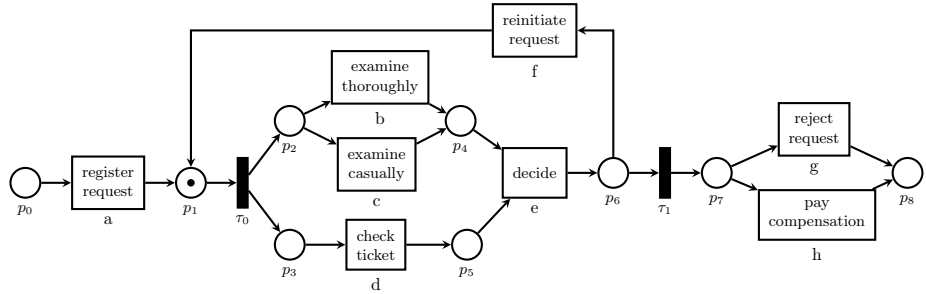
\begin{figure}[htp]
    \centering
    \scalebox{0.66}{
\begin{tikzpicture}[
    node distance=6mm,very thick,align=center,
    every transition/.style={minimum height=1cm,inner sep=2mm,font=\small},
    every place/.style={minimum size=6mm,font=\small}]
\node[place,label=below:{\small$p_0$}] (p0) {};
\node[transition,right=of p0,label=below:{a}] (t0) {register\\request};
\node[place,right=of t0,label=below:{\small$p_1$},tokens=1] (p1) {};
\node[transition,right=of p1,label=below:{\small$\tau_0$},minimum width=2mm, inner sep=0,fill=black] (tau0) {};
\node[place,above right=5mm and 6mm of tau0,label=below:{\small$p_2$}] (p2) {};
\node[place,below right=5mm and 6mm of tau0,label=below:{\small$p_3$}] (p3) {};
\node[transition,above right=1mm and 6mm of p2,label=below:{b}] (t1) {examine\\thoroughly};
\node[transition,below right=1mm and 7.2mm of p2,label=below:{c}] (t2) {examine\\casually};
\node[place,right=28mm of p2,label=below:{\small$p_4$}] (p4) {};
\node[transition,below right=3mm and6mm of p4,label=below:{e}] (t3) {decide};
\node[transition,right=8mm of p3,label=below:{d}] (t4) {check\\ticket};
\node[place,right=9mm of t4,label=below:{\small$p_5$}] (p5) {};
\node[transition,above= 18mm of t3,label=below:{f}] (t5) {reinitiate\\request};
\node[place,right=of t3,label=below:{\small$p_6$}] (p6) {};
\node[transition,right=of p6,label=below:{\small$\tau_1$},minimum width=2mm, inner sep=0,fill=black] (tau1) {};
\node[place,right=of tau1,label=below:{\small$p_7$}] (p7) {};
\node[transition,above right=1mm and 1cm of p7,label=below:{g}] (t6) {reject\\request};
\node[transition,below right=1mm and 6mm of p7,label=below:{h}] (t7) {pay\\compensation};
\node[place,right=32mm of p7,label=below:{\small$p_8$}] (p8) {};

\draw[-{stealth}] (p0) edge (t0) (t0) edge (p1) (p1) edge (tau0) (tau0) edge (p2) (tau0) edge (p3) (p2) edge (t1) (p2) edge (t2) (t1) edge (p4) (t2) edge (p4) (p3) edge (t4) (t4) edge (p5) (p5) edge (t3) (p4) edge (t3) (t3) edge (p6) (p6) edge (tau1) (tau1) edge (p7) (p7) edge (t6) (p7) edge (t7) (t6) edge (p8) (t7) edge (p8);
\draw[-{stealth}] (p6) |- (t5);
\draw[-{stealth}] (t5) -| (p1);
\end{tikzpicture}
}
    \caption{\textcolor{black}{Accepting Petri net}}
    \label{fig:running-example}
\end{figure}

The operational semantics of an accepting Petri net is defined in terms of markings. A marking $M$ \emph{enables} a transition $t$ if $\forall p \in \pre{t} : M(p) > 0$, denoted as $(N,M)[t\rangle$. Moreover, the occurrence of $t$ leads to a new marking $M^\prime$, with $M^\prime(p) = M(p) - 1$ if $p \in \pre{t} \setminus \post{t}$, $M^\prime(p) = M(p) + 1$ if $p \in \post{t} \setminus \pre{t}$, and $M^\prime(p) = M(p)$ otherwise. We use $M\larrow{t}M^\prime$ to denote the occurrence of $t$. The marking $M_n$ is said to be reachable from $M$, shorthanded as $M\larrow{\sigma}M_n$, if there exists a sequence of transitions $\sigma= \langle t_1 t_2 \dots t_n \rangle$ such that $M\larrow{t_1}M_1\larrow{t_2}\dots\larrow{t_n}M_n$. A marking $M$ where no transition is enabled is called \emph{terminal}, i.e., $\nexists ~t \in T ~|~ (N,M')[t\rangle$. The set of all the markings reachable from a marking $M$ is denoted as $[M\rangle$.
% --- Firing sequence ...
A sequence of transitions from an initial marking $M$ to a final marking $M'$ is called a run. 
% -----
The set of all markings reachable from a marking $M$ is denoted $[ M \rangle$. 
A marking $M$ of a net is \emph{n-bounded} if $M(p) \leq n$ for every place $p$. An accepting Petri net $\mathcal{N}$ is said \emph{n-bounded} if all its reachable markings are \emph{n-bounded}. It is called \emph{safe} if it is \emph{1-bounded}. 

% ---- Multiset notation
We will often represent marking using a multiset-like notation. For instance, $M = \{p_2^{(1)}, p_6^{(1)}\}$ would be used to state the fact that places $p_2$ and $p_6$ hold one token each. Every other place in the accepting Petri net is assumed to hold none.

Figure~\ref{fig:running-example} displays an accepting Petri net that will be used as the running example. This Petri net has a net where the initial marking is $M_0 = \{p_0^{(1)}\}$, and the final marking is $M_f = \{p_8^{(1)}\}$. Note that the marking represented in Figure~\ref{fig:running-example} can be the marking $M'$ reachable from $M_0$, i.e., $M_0\larrow{a}M^\prime$.

% \textcolor{red}{I think we should only consider sound, safe Workflow nets ... \textcolor{blue}{@Abel} what do you think about this assumption? If agree then we would have to define workflow nets and soundness criteron.}

% In the following, we assume that all the Petri nets are sound and safe. 

\subsubsection{Incidence matrix and Marking equation}

An alternative approach to reasoning about the dynamics of net systems is using operations on matrices~\cite{FreeChoiceNets}. The structure of the net itself is represented by an incidence matrix, markings as column vectors and firing sequences as row vectors.

\begin{definition}[Incidence matrix of a net]\\
Let $N$ be the net $(P,T,F)$. Hence, the incidence matrix $\mathbf{N}: (P\times T) \to \{-1,0,1\}$ of $N$ is defined as:
\[
    \mathbf{N}(p,t) = \left\{
    \begin{array}{rm{8mm}l}
    -1 & & \text{\emph{if $(p,t) \in F$ and  $(t,p) \notin F$}},\\
    1 & & \text{\emph{if $(t,p) \in F$ and  $(p,t) \notin F$}, \emph{and}}\\
    0 & & \text{\emph{otherwise}}
    \end{array}
    \right.
\]
\end{definition}

The incidence matrix of the net in Figure~\ref{fig:running-example} is presented in Figure~\ref{fig:sample-marking-equation}, with the name $\mathbf{N}$. As customary, the names of the matrices and vectors are typed in bold font. It can be noted that the rows are associated with places in the net, which, for convenience, we ordered lexicographically. The columns, on the other hand, correspond to transitions. Also, for convenience, we first enumerated the visible transitions (in lexicographical order), leaving all silent transitions in the last columns.

In this context, markings are also represented as column vectors. In Figure~\ref{fig:sample-marking-equation}, on the left-hand side, we have the marking $M = \{p_1^{(1)}\}$, which in matrix notation would be represented as $\mathbf{M} = \left[\text{0 1 0 0 0 0 0 0 0}\right]^T$. In the following, we will consider both representations as equivalent.

We also represent sequences of firing transitions, commonly referred to as Parikh vectors, as row vectors. For instance, the firing sequence $\left<\tau_0, d\right>$ is represented as the firing sequence $\boldsymbol{\sigma}=\left[\text{0 0 0 1 0 0 0 0 1 0}\right]$. For convenience, the position of the transitions follows the alphabetical order of their names. The value at each position corresponds to the number of times a transition must be fired. The example of firing sequence above is also included in Figure~\ref{fig:sample-marking-equation}, although transposed. %Please note that in this work, we will restrict ourselves to firing sequences where each transition to fire does it only once. 

%$M_i \left|\tau_0\right> M_{i+1} \left|d\right> M_{i+2}$

\begin{figure}[ht]
    \centering
\begin{tikzpicture}
\node at (0,0) [label=below:{$\mathbf{M}$}] {$
\begin{bNiceMatrix}%
[first-row,first-col]
\\
p_0 & 0\\
p_1 & 1\\p_2 & 0\\p_3 & 0\\p_4 & 0\\p_5 & 0\\p_6 & 0\\p_7 & 0\\p_8 & 0\\
\end{bNiceMatrix}
$};
\node at (0.7,0) {+};
\node at (4.4,0) [label=below:{$\mathbf{N}$}] {
$
\begin{bNiceMatrix}%
    [first-row,first-col,columns-width=5mm]
 & \text{a} & \text{b} & \text{c} & \text{d} & \text{e} & \text{f} & \text{g} & \text{h} & \tau_0 & \tau_1\\
p_0\phantom{.} & -1 & 0 & 0 & 0 & 0 & 0 & 0 & 0 & 0 & 0\\
p_1\phantom{.} &  1 & 0 & 0 & 0 & 0 & 1 & 0 & 0 & -1 & 0\\
p_2\phantom{.} &  0 & -1 & -1 & 0 & 0 & 0 & 0 & 0 & 1 & 0\\
p_3\phantom{.} &  0 & 0 & 0 & -1 & 0 & 0 & 0 & 0 & 1 & 0\\
p_4\phantom{.} &  0 & 1 & 1 & 0 & -1 & 0 & 0 & 0 & 0 & 0\\
p_5\phantom{.} &  0 & 0 & 0 & 1 & -1 & 0 & 0 & 0 & 0 & 0\\
p_6\phantom{.} &  0 & 0 & 0 & 0 & 1 & -1 & 0 & 0 & 0 & -1\\
p_7\phantom{.} &  0 & 0 & 0 & 0 & 0 & 0 & -1 & -1 & 0 & 1\\
p_8\phantom{.} &  0 & 0 & 0 & 0 & 0 & 0 & 1 & 1 & 0 & 0\\
\end{bNiceMatrix}
$
};
\node at (8,0) {$\cdot$};
\node at (8.7,0) [label=below:{$\boldsymbol{\sigma}^T$}]{$
\begin{bNiceMatrix}[last-col]
0 & a\\
0 & b\\0 & c\\1 & d\\0 & e\\0 & f\\0 & g\\0 & h\\1 & \tau_0\\0 & \tau_1\\
\end{bNiceMatrix}
$};
\node at (9.5,0) {=};
\node at (10.2,0) [label=below:{$ \mathbf{M}^\prime $}] {$
\begin{bNiceMatrix}[first-row]
\\
0\\
0\\1\\0\\0\\1\\0\\0\\0\\
\end{bNiceMatrix}
$};
\end{tikzpicture}
    \caption{\textcolor{black}{Marking equation, where $\mathbf{M}$ and $\mathbf{N}$ represent the marking and incidence matrix of Fig. ~\ref{fig:running-example} and $\boldsymbol{\sigma}^T$ corresponds to the firing sequence $\left<\tau_0, d\right>$}}
    \label{fig:sample-marking-equation}
\end{figure}
% \todo[inline]{Creo que las files p3 y p4 de N están al revés}
With the above elements, it is possible to reformulate the execution dynamics of a Petri net. To that end, we can use the \emph{Marking equation}~\cite{FreeChoiceNets}.

\begin{definition}[Marking equation]
Let $M\larrow{t}M^\prime$ be a finite occurrence sequence of a net $N$. Then, $M'$ can be obtained with the following \emph{marking equation}:
$$
\mathbf{M}^\prime = \mathbf{M} + \mathbf{N} \cdot \boldsymbol{\sigma}^T
$$

\end{definition}

Figure~\ref{fig:sample-marking-equation} shows the full example of the firing of a pair of transitions from a given marking. Assuming we start at the marking $M=\{p_1^{(1)}\}$, and that we want to compute the marking reached after firing the following sequence $\sigma = \left<\tau_0, d\right>$. It can be seen that, for the above conditions, we reach the marking $M'=\{p_2^{(1)},p_5^{(1)}\}$, as expected. Note that firing a disabled transition will result in a marking with negative values. However, in this work, we check first if the transition to be fired is enabled to avoid this problem.

\subsection{Event logs}

An event log consists of a set of traces, each recording the execution of a process instance. In turn, a trace is a totally ordered sequence of events. Each event corresponds to a transition in the life cycle of a task (e.g. a task became enabled, the execution of a task started or completed, etc.). Event logs are formally defined as follows:

\begin{definition}[Event log]
Let $\mathcal{L} \subseteq \mathcal{A}$ be a set of activity names. A trace $\theta$ is a sequence drawn from the Kleene closure of $\mathcal{L}$, i.e., $\theta \in \mathcal{L}^*$. An event log is a multiset of traces, i.e., $L \in \mathcal{B}(\mathcal{L}^*)$. With abuse of notation, if $\theta \in L$ we will denote the multiplicity of $\theta$ as $L(\theta)$.
\end{definition}

Since a process may have multiple executions resulting in identical traces, an event log is defined as a multiset. In some contexts, each element in the multiset is referred to as a trace variant. When a single representative per trace variant is needed, we define the function $set : \mathcal{B}(\mathcal{L}^*) \to \mathcal{L}^*$ as $\{\theta\ |\ \theta \in L \land L(\theta) > 0\}$. Therefore, if $L$ is an event log of a process, then $set(L)$ is the corresponding set of trace representatives.

\subsection{Token-based replay conformance checking}
\label{subsec:token-based-replay}

Token-based replay is inspired on the token game which is often used for discussing the execution of a Petri net. Given a trace, token-based replay checks whether the net, starting from the initial marking, can fire a sequence of transitions matching the events and their order in the trace, so that the final marking is eventually reached. Note that token-based replay may require firing a number of $\tau$ transitions.

For instance, consider the trace $\left<a,b,d,e,h\right>$ and the Petri net in Figure~\ref{fig:running-example}. Token-based replay starts with marking $M_0 = \{p_0^{(1)}\}$, and induces the following firing sequence:\\
{%\small
\[
\{p_0^{(1)}\}\ \left|\textcolor{red}{a}\right>\ \{p_1^{(1)}\}\ \left|\textcolor{red}{\tau_0, b}\right>\ \{p_3^{(1)}, p_4^{(1)}\}\ \left|\textcolor{red}{d}\right>\ \{p_4^{(1)}, p_5^{(1)}\}\ \left|\textcolor{red}{e}\right>\ \{p_6^{(1)}\}\ \left|\textcolor{red}{\tau_1, h}\right>\ \{p_8^{(1)}\}
\]}\\
\noindent
which reaches the net's final marking. It is worth noting that the whole token game used a total of 8 tokens. We say that token-based replay \emph{produced} 8 tokens, all of which were later \emph{consumed}. We say that a trace \emph{fits} the behavior specified by a net if there exists a sequence of transition firings that reach the final state.

In the occurrence of a discrepancy, on the other hand, token-based replay gets stuck: the next event in the trace matches a transition that is not enabled by the current marking. To cope with this situation, token-based replay inserts the \emph{missing} tokens in the preset of the transition that is stuck to let the token game proceed. For instance, the trace $\left<a,b,e,h\right>$ would induce the following firing sequence:

\begin{tabular}{l}
\hspace{-6mm}$\{p_0^{(1)}\}\ \left|\textcolor{red}{a}\right>\ \{p_1^{(1)}\}\ \left|\textcolor{red}{\tau_0, b}\right>\ \{p_3^{(1)}, p_4^{(1)}\}$ \textcolor{red}{\XSolidBrush} \hspace{2mm} $e$ is not enabled, then add tokens to $\bullet e$\\ 
\hspace{5mm}\textcolor{green}{\ArrowBoldDownRight}$
\{p_3^{(1)}, p_4^{(1)}\} \cup\highlight{yellow!40}{\{p_4^{(1)},p_5^{(1)}\}}\ \left|\textcolor{red}{e}\right>\ \{p_3^{(1)}, p_4^{(1)},p_6^{(1)}\}\ \left|\textcolor{red}{\tau_1, h}\right>\ \{\highlight{orange!40}{p_3^{(1)}, p_4^{(1)}},p_8^{(1)}\}$
\end{tabular}

\noindent
Despite the final marking is reached, i.e., $\{p_8^{(1)}\}$, other tokens are left behind, i.e., $\{p_3^{(1)},p_4^{(1)}\}$. Hence, in the case of discrepancies, \emph{missing} tokens need to be added, and some will be \emph{remaining} when the final marking is reached. In the above example, token-based replay adds 2 missing tokens  (highlighted in yellow), and ends up with 2 remaining tokens (highlighted in orange). 

In order to quantify the presence of the discrepancies, the notion of \emph{fitness} has been introduced~\cite{Aalst16}. A trace that can be fully replayed without the need to insert missing tokens is reported with a fitness of 1. The occurrence of discrepancies decreases the value of fitness. In the following, we include the definition of token-based replay fitness as introduced in~\cite{Rozinat08}. 
% \todo{I put a missing citation, is the Berti2021?}

\begin{definition}[Token-based replay fitness]
    Let $\mathcal{N} = (N,M_0, M_f)$ be an accepting Petri net, $L$ be an event log and $\theta$ a trace of $L$. Moreover, let $c, p, m$ and $r$ be the number of consumed, produced, missing and remaining tokens induced by the replaying a $\theta$ against the net $\mathcal{N}$. Then the fitness value of trace $\theta$ w.r.t. $\mathcal{N}$ is computed as:
    $$
    f(\mathcal{N},\theta) = 
        \frac{1}{2}\left(1 - \frac{m}{c}\right)
        + \frac{1}{2}\left(1 - \frac{r}{p}\right)$$
\end{definition}

The notion of log's token-based replay fitness is also defined, as a way to aggregate a value of fitness for the entire event log, and is straightforwardly computed using $c,\ p,\ m$ and $r$. 
%\textcolor{red}{However, in our setting, it is unlikely that a single stakeholder has access to the entire event log.}\todo{I do not understand this.}

Since some missing tokens can be inserted during replay, it might be that some places end up with more than one token, even if the underlying Petri net is safe. This is known as token flooding, a situation that can arise with this technique.~\cite{Rozinat08} describes rules that can be applied to cope with this situation.

\subsection{Homomorphic encryption}

Homomorphic encryption refers to a technique of public-key cryptography that enables computation on encrypted data. The latter implies that a party can manipulate encrypted data, i.e., with arithmetic operations, without the need of decrypting the data, hence keeping secret the underlying information. The ideas behind homomorphic encryption were first discussed in the late 70s~\cite{Rivest78}.
% , and came to life only in the 90s~\cite{}. 
Initially, only one operation over integers was supported, i.e., either addition or multiplication, such that it is referred to as Partial Homomorphic Encryption. In 2009, \cite{Gentry09} %\todo{Missing citation} 
described the first implementation of what is now referred to as ``Fully Homomorphic Encryption'', because it supports both additions and multiplications. However, the implementation, and some that followed, were not practical for real-world applications due to its inherent high computational overheads. Recent advancements in the field have led to faster implementations, which include support to a wider repertoire of operations on booleans, integers and even floating point numbers. We highlight the fact that companies have emerged proposing machine learning techniques for encrypted data using fully homomorphic encryption (e.g., \cite{FrerySBMKCM23,MonteroFKBS24}).

In the context of this work, we explore the use of homomorphic encryption to implement a secure conformance checking technique. The inherent challenge consists in reformulating token-based replay using arithmetic operations over integers. In the end, our algorithm required a richer set of operations, which we found available in Zama's Concrete~\cite{Concrete}, an open source platform for Python. Note that in this work we focus on the description of our adaptation of the token-based replay, which we have successfully implemented using Zama's Concrete. We do not go deeper in the details on how homomorphic encryption is implemented as it is beyond de scope of our work.

Zama's Concrete is packaged as a Python package, which provides tools to define functions that manipulate homomorphically encrypted data, which are later compiled and linked with some native libraries. All those native libraries are implemented in Rust. However, Concrete allows us to specify all the homomorphic operations using Python syntax. Interestingly, the package offers a wrapper for NumPy, such that some vector and matrix operations can be directly used over encrypted data. Our implementation benefits from the latter and uses: multiplications, additions and subtractions on matrices and vectors of integers, division of integers, as well as the function to select the maximum and minimum element of a pair of values, among others. Please note that some of these operations are not available in other frameworks. However, there is a large amount of work that describes approaches to implement all the operations required by our algorithm on top of other platforms (e.g., \cite{ChakrabortyZ22,Veugen14}).

\section{Approach}
\label{sec:approach}

\begin{figure}
    \centering
    \includegraphics[width=0.85\linewidth]{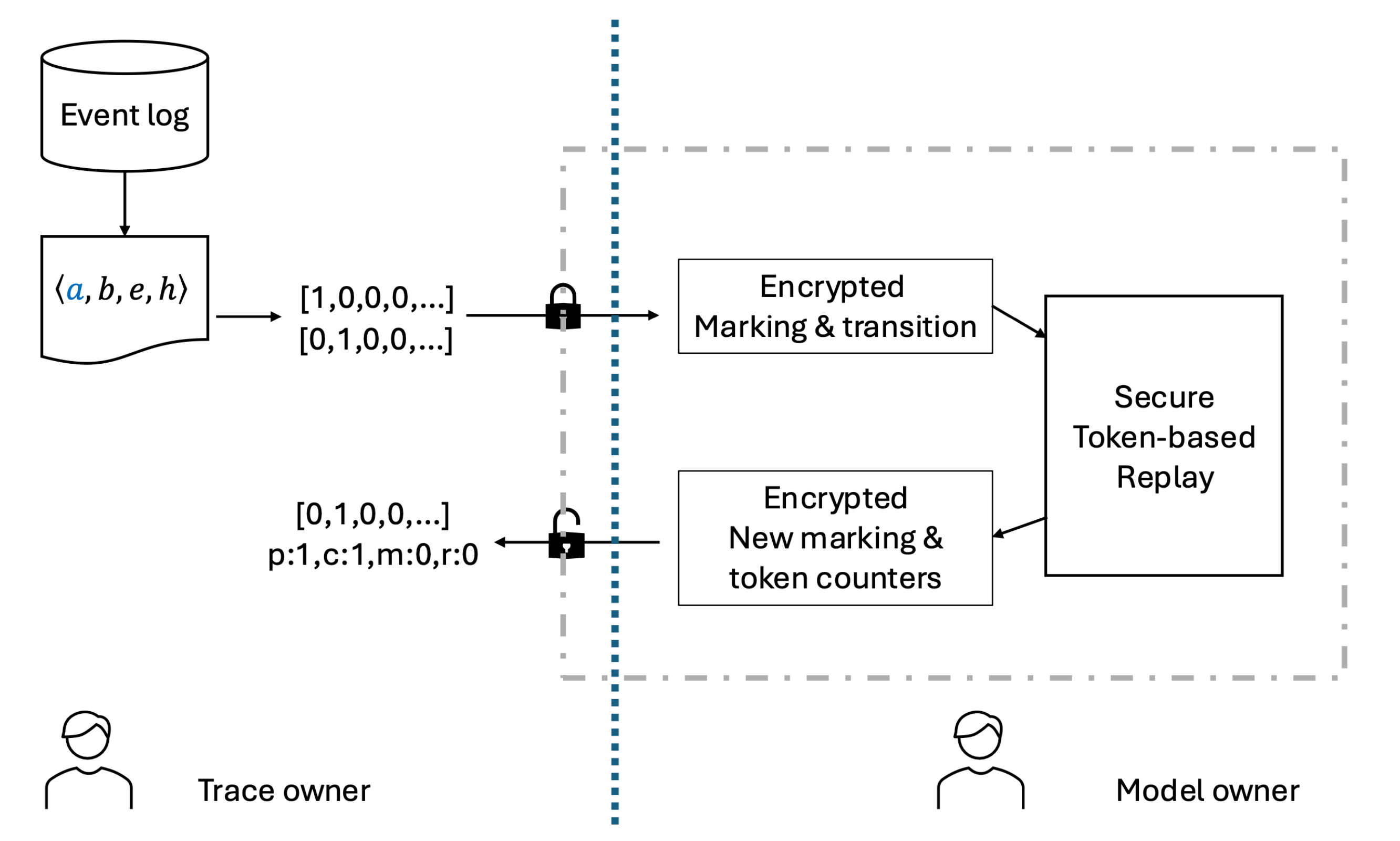}
    \caption{Block diagram illustrating the approach and system architecture}
    \label{fig:enter-label}
\end{figure}

As stated before, our aim is to have an algorithm that allows us to implement the token-based replay in a setting where the model and the trace are held by different parties. The trace owner will iteratively verify the conformance of its trace, by encoding the execution of a trace using a vector of integers, where one element is set to one while the other to zeroes. Moreover, the same party uses a vector, representing the marking, initialized with all zeroes. Both vectors are encrypted using a private key and sent to the model's owner. The latter receives the vectors and runs one step of the token-based replay locally, generating the new marking and the statistics to compute the trace's fitness. The architecture of the system allowing the interactions between the trace and the model owners is shown in Figure~\ref{fig:enter-label}.

Clearly, the system reflects a client/server architecture. The trace's owner plays the role of the client, which iteratively requests the evaluation of one step of the token-based replay. The model's owner waits for requests and processes them accordingly. Since the requests and responses are encrypted (the information is never decrypted), the model's owner does not get to know the trace.

In the following, we define our algorithm for token-based replay using matrix and vector operations, which has been implemented using homomorphic encryption. The presentation is organized in three stages. First, we discuss how to handle fitting traces. Second, we describe how to support the insertion of missing tokens to handle unfitting traces. Third, we explain how to compute the counters $p,\ c,\ m$ and $r$ over encrypted intermediate information. Finally, we present the full algorithm and discuss some aspects related to homomorphic encryption.

\subsection{Fitting traces}

Let us first consider the case where a trace can be fully replayed by the underlying Petri net. To illustrate the concepts, let us assume that we want to check the conformance of the trace $\left<a,d,b,e,h\right>$ in the accepting net displayed in Figure~\ref{fig:running-example}.

Token-based replay starts by considering the initial making $M_0 = \{p_0^{(1)}\}$ and proceeds trying to execute every event in the trace, one at a time. That means that we will try to execute first the event $a$, which happens to be enabled, i.e., $\bullet{a} \subseteq set(M_0)$. After firing $a$, we reach the marking $M_1 = \{p_1^{(1)}\}$. We now will try to fire the second event, transition $d$. In this case, transition $d$ is not enabled, but it will become enabled if silent transition $\tau_0$ were fired. Due to the existence of $\tau_0$, there are two possible markings that enable the replaying of transition $d$: 
\begin{inparaenum}[i)]
    \item with marking $M_1 = \{p_1^{(1)}\}$, inducing the firing of the sequence $\left<\tau_0,d \right>$, and 
    \item with marking $M_1' = \{p_3^{(1)}\}$, inducing the firing of the sequence $\left<d \right>$. 
\end{inparaenum}

%The challenge here is to determine the firing sequence to be used by considering the current marking and the next event from the log to be processed, using matrix operations. 
The core of the approach is to gather all possible scenarios (marking and transition to be executed) in a matrix that will be manipulated algebraically. For our running example, the aforementioned matrix is shown in Figure~\ref{fig:enablements}. In the example, we want to execute transition \textcolor{black}{$d$} 
%b
for which we have two scenarios shown inside a red box -- rows 5 and 6 in Figure~\ref{fig:enablements} (the row numbers are shown at the left-hand side of the matrix). In row 5, we see that we need a token in place $p_1$ and will require the firing of transition $\tau_0$ -- the elements are highlighted elements in rows 5 and 6 to help the reader follow the explanation. Similarly, row 6 specifies that transition $d$ can be fired immediately if place $p_3$ holds a token. 

% ========================
\setcounter{MaxMatrixCols}{21}

\begin{figure}
    \centering
\scalebox{1}{
    \small
\begin{tikzpicture}
\node at (0,0) {$
\begin{bNiceMatrix}%
    [first-row,first-col]
\CodeBefore
 \tikz \draw[color=red,thick]  (6-|1) -- (6-|22) -| (8-|22) -- (8-|1) |- cycle;
\Body
 & p_0 & p_1 & p_2 & p_3 & p_4 & p_5 & p_6 & p_7 & p_8 & & a & b & c & d & e & f & g & h & & \tau_0 & \tau_1\\
\text{\scriptsize 0} & \phantom{1}1\phantom{1} & \phantom{1}0\phantom{1} & 
\phantom{1}0\phantom{1} & \phantom{1}0\phantom{1} &  \phantom{1}0\phantom{1} &  \phantom{1}0\phantom{1} &  
\phantom{1}0\phantom{1} & \phantom{1}0\phantom{1} &
\phantom{1}0\phantom{1} & & \phantom{1}1\phantom{1} & \phantom{1}0\phantom{1} & \phantom{1}0\phantom{1} & \phantom{1}0\phantom{1} &
\phantom{1}0\phantom{1} & \phantom{1}0\phantom{1} & \phantom{1}0\phantom{1} &
\phantom{1}0\phantom{1} &  & \phantom{1}0\phantom{1} & \phantom{1}0\phantom{1}\\
% ---- b
\text{\scriptsize 1} & 0 & 1 & 0 & 0 & 0 & 0 & 0 & 0 & 0 & & 0 & 1 & 0 & 0 & 0 & 0 & 0 & 0 & & 1 & 0 \\
\text{\scriptsize 2} & 0 & 0 & 1 & 0 & 0 & 0 & 0 & 0 & 0 & & 0 & 1 & 0 & 0 & 0 & 0 & 0 & 0 & & 0 & 0 \\
% ---- c
\text{\scriptsize 3} & 0 & 1 & 0 & 0 & 0 & 0 & 0 & 0 & 0 & & 0 & 0 & 1 & 0 & 0 & 0 & 0 & 0 & & 1 & 0 \\
\text{\scriptsize 4} & 0 & 0 & 1 & 0 & 0 & 0 & 0 & 0 & 0 & & 0 & 0 & 1 & 0 & 0 & 0 & 0 & 0 & & 0 & 0 \\
% ---- d
\text{\scriptsize 5} & 0 & \cellcolor{yellow!50}1 & 0 & 0 & 0 & 0 & 0 & 0 & 0 & & 0 & 0 & 0 & \cellcolor{yellow!50}1 & 0 & 0 & 0 & 0 & & \cellcolor{yellow!50}1 & 0 \\
\text{\scriptsize 6} & 0 & 0 & 0 & \cellcolor{yellow!50}1 & 0 & 0 & 0 & 0 & 0 & & 0 & 0 & 0 & \cellcolor{yellow!50}1 & 0 & 0 & 0 & 0 & & 0 & 0 \\
% ---- e
\text{\scriptsize 7} & 0 & 0 & 0 & 0 & 1 & 1 & 0 & 0 & 0 & & 0 & 0 & 0 & 0 & 1 & 0 & 0 & 0 & & 0 & 0 \\
% ---- f
\text{\scriptsize 8} & 0 & 0 & 0 & 0 & 0 & 0 & 1 & 0 & 0 & & 0 & 0 & 0 & 0 & 0 & 1 & 0 & 0 & & 0 & 0 \\
% ---- g
\text{\scriptsize 9} & 0 & 0 & 0 & 0 & 0 & 0 & 1 & 0 & 0 & & 0 & 0 & 0 & 0 & 0 & 0 & 1 & 0 & & 0 & 1 \\
% ---- g
\text{\scriptsize 10} & 0 & 0 & 0 & 0 & 0 & 0 & 1 & 0 & 0 & & 0 & 0 & 0 & 0 & 0 & 0 & 0 & 1 & & 0 & 1 \\
\CodeAfter
\UnderBrace[yshift=1mm,color=blue]{11-1}{11-18}{\text{Enablements }(\mathcal{E})}
\OverBrace[yshift=1mm,color=blue]{0-11}{0-21}{\text{Firing sequences }(\mathcal{S})}
\end{bNiceMatrix}
$};
\end{tikzpicture}
}
 \hspace{3mm}
    \caption{Matrix of dynamics of Petri net in Figure~\ref{fig:running-example}}
    \label{fig:enablements}
\end{figure}

% ========================

It can be easily checked that the matrix in Figure~\ref{fig:enablements} contains information to determine whether a transition can be fired considering a given marking. In some cases, firing that transition requires the firing of an additional set of invisible transitions. For convenience, the matrix is divided into two sub-matrices that we call \emph{enablements}, denoted as $\mathcal{E}$, and \emph{firing sequences}, denoted as $\mathcal{S}$.

% \todo[inline,color=red]{I have to say that I get lost in this following explanation. }
During replay, the next event to be fired is encoded as a vector, that is, the Parikh vector of the corresponding transition, which we will denote $\mathbf{t}$. We then determine, given the current marking, which scenario applies by using some matrix operations. The marking and Parikh vector are concatenated to form a column vector and we compute the product:
$\mathcal{E}\cdot\left[\mathbf{M}^T \mathbf{t} \right]^T$.
In our example, where we are trying to fire transition $d$ with marking $\{p_1^{(1)}\}$, the above corresponds to multiplying the matrix $\mathcal{E}$ in Figure~\ref{fig:enablements} with the column vector [\textcolor{red}{0 1 0 0 0 0 0 0 0} \textcolor{blue}{0 0 0 1 0 0 0 0}]$^T$ 
% \todo{I think the Parikh vector of d is wrong}
(the marking is shown in red, and the Parikh vector in blue). Note that the Parikh vector does not include the invisible transitions, i.e., in our setting, the owner of the trace is not aware of the structure of the net. The matrix product results in the column vector [0 1 0 1 0 2 1 0 0 0 0]$^T$. Let us call the latter $\mathbf{tmp}$. Note that the resulting vector contains several entries with a value $> 0$ whenever the marking indicates that $p_1$ has a token (i.e., rows 1, 3, 5) or whenever the transition to be fired is $b$ (i.e., rows 5, 6). It can be easily seen that we are interested in the case where both conditions hold, marking and transition match. In order to check the marking/transition condition, we compute the expected number per row. In most of the cases in this example, we will expect the value to be 2. The only exception is in row 7, where the marking must have tokens in two places hence, we expect to have a value of 3. We then define an auxiliary vector that we call \emph{divisors}, denoted $\boldsymbol{\Delta}$, which can be computed by summing the entries on each row of $\mathcal{E}$. In our example, $\boldsymbol{\Delta}$ = [2 2 2 2 2 2 2 3 2 2 2]$^T$. Let us define $\div$ as the element-wise integer division between two vectors. Hence, $\mathbf{tmp} \div \boldsymbol{\Delta}$ will result in the column vector [0 0 0 0 0 1 0 0 0 0 0]$^T$, which we will call the \emph{selector}, denoted $\mathbf{sel}$. Note that if a trace is fitting (i.e., there is a firing sequence in the net for the trace), this vector will always contain a single element with value 1, and all the other values will be 0. As the final step, we will multiply each row in the $\mathcal{S}$ to the corresponding element in the same row (recall it is a column vector), and sum up all the resulting rows. It can easily be seen that only the row in position 5, where $\mathbf{sel}$ has a value of 1, will be kept and the other rows will be discarded. That allows us to conclude that the firing sequence (Parikh vector) to be used is [0 0 0 1 0 0 0 0 1 0], which corresponds to the sequence $\left|\tau_0,d\right>$, and that we will denote $\boldsymbol{\sigma}$.
% \todo[inline]{I'm not sure if I understood correctly. Aren't we trying to fire transition $d$? If that's the case, there are a couple of references to transition $b$ in the text (including the final firing sequence).
% Also, is $\boldsymbol{\Delta}$ correct? Wouldn't the 1 be in position 5?
%%%}

What we have done so far is the following:
\begin{equation}
\label{eq:selector}
    \mathbf{sel} \gets \mathcal{E} \cdot \left[\begin{matrix} \mathbf{M}\\ \mathbf{t}^T \end{matrix}\right] \div \boldsymbol{\Delta}
\end{equation}
\begin{equation}
    \boldsymbol{\sigma} \gets \sum_{i = 0}^{|\mathbf{sel}|-1}\mathcal{S}[i]\cdot\mathbf{sel}[i]
\end{equation}

Having defined which firing sequence must be executed, we can simply apply the marking equation:
\begin{equation}
\label{eq:marking1}
\mathbf{M}^\prime \gets \mathbf{M} + \mathbf{N}\cdot \boldsymbol{\sigma}^T
\end{equation}

and repeat with all the events in the trace to complete the token-based replay.

It is worth noting, as it will become evident, in some cases the marking may have some places with more than one token despite the net being safe. To cope with this situation, we need to preprocess the marking to ensure that all the elements are zeroes or ones. Recall a marking can be seen as a multiset, therefore the preprocessing step corresponds to computing the corresponding set, that means computing $\hat{M} = set(M)$. Using matrix operations we resort to a vector function $min$, which compares to vectors (column or row vectors) element-wise and selects the minimum value thereof. Let $\mathbf{0}$ denote a vector with only zeroes as elements and $\mathbf{1}$ a vector with only ones. Hence, we can specify $\hat{M} = set(M)$ as:
\begin{equation}
    \mathbf{\hat{M}} \gets \text{min}(\mathbf{M}, \mathbf{1})
\end{equation}
\noindent
Therefore, we only need to replace $\mathbf{M}$ by $\mathbf{\hat{M}}$ in the computation of $\mathbf{sel}$, that is, in Equation~\ref{eq:selector}.

\subsection{Handling unfitting traces}

In Subsection~\ref{subsec:token-based-replay}, we analyzed the replay of trace $\left<a,b,e,h\right>$. After executing transition $a$ and transition $b$, we reached the marking $\{p_3^{(1)}, p_4^{(1)}\}$, which as discussed before does not enable transition $e$. The palliative was to insert the \emph{missing} tokens to the marking to allow the replay to continue.

To implement the above, we precompute a matrix with all the presets of the transitions on the net, which we will refer to as $\mathcal{P}$. On the other hand, recall that we use row vector $\mathbf{t}$ to specify which transition we want to fire. The corresponding entry will have a 1 and all the other entries zeroes. Therefore, we use $\mathbf{t}$ as the selector. Therefore, we can determine the preset, i.e., $\boldsymbol{\pi}$, of the transition to be fired as follows:
\begin{equation}
    \boldsymbol{\pi} \gets \sum_{i = 0}^{|\mathbf{t}|-1}\mathcal{P}[i]\cdot\mathbf{t}[i]
\end{equation}
\noindent
Note that the above implies multiplying each row of $\mathcal{P}$ times the scalar in $\mathbf{t}[i]$. The marking equation is adapted to consider the missing tokens as shown below:
\begin{equation}
\label{eq:marking2}
\mathbf{M}^\prime \gets \mathbf{M} + \boldsymbol{\pi}^T + \mathbf{N}\cdot \mathbf{t}^T
\end{equation}

At this point, we have two different equations to compute the next marking. For security reasons, in systems using homomorphic encryption, the use of conditional statements is avoided. Its use could potentially leak information, if one tracks the execution of the code. To cope with this limitation, we use the following approach. First, we note that $\mathbf{sel}$ will have one single element with value of $1$,
that is, when there exists a firing sequence for a given transition that can be enabled with the current marking.
On the other hand, if the replay reaches a point where some missing tokens need to be inserted to proceed, all the elements in $\mathbf{sel}$ will be zero. After this observation we can see that if we sum up all the elements of {\bf sel}, we will have either 1 (there exists a firing sequence) or 0 (we need to add some missing tokens to proceed), and henceforth we can combine Equations~\ref{eq:marking1} and \ref{eq:marking2} as follows:
\begin{equation}
\mathbf{M'} \gets \left(\mathbf{M} + \mathbf{N} \cdot \boldsymbol{\sigma}^T \right)\cdot \text{sum}(\mathbf{sel}) + \left(\mathbf{M} + \boldsymbol{\pi}^T + \mathbf{N} \cdot \mathbf{t}^T\right)\cdot \left(1-\text{sum}(\mathbf{sel})\right)
\end{equation}

Note that we use a function sum that adds the elements in a vector. Since sum($\mathbf{sel}$) is either 0 or 1, $1 - \text{sum}(\mathbf{sel})$ will \emph{invert} the value. Hence, this value serves as a selector between the two marking equations.

\subsection{Computing the counters $p,\ c,\ m,$ and $r$}

We turn our attention to discussing how to compute the token counters using matrix operations. Let us start with the number of missing tokens $m$. Its computation is straightforward, as it corresponds to the number of places in the preset of the transition being fired. This can be done as shown in the following equation:
\begin{equation}
m \gets \text{sum}(\boldsymbol{\pi}) \cdot (1 - \text{sum}(\mathbf{sel}))
\end{equation}
\noindent
The expression $1 - \text{sum}(\mathbf{sel})$ is used again to decide whether the value is to be taken into account or not.

Counters $p$ and $c$ are derived from the current marking and the marking after firing the transition at hand. From our running example, the marking ${M} =\{p_1^{(1)}\}$ turns into ${M}^\prime=\{p_3^{(1)},p_4^{(1)}\}$ after executing the sequence $\left|\tau_0,b \right>$. Since markings are multisets, it will become evident that computing $p$ and $c$ relies on a multiset difference. In the case of fitting traces, $c = |M \setminus M^\prime|$ and $p = |M \setminus M^\prime|$. The multiset difference $M \setminus M'$ using the vector representation is implemented as max$((\mathbf{M} - \mathbf{M}^\prime), \mathbf{0})$, as in Definition~\ref{def:multiset}. To illustrate the idea, consider the vector operation [1 0 0] - [0 1 1], which results in [1 -1 -1]. The use of the function max is required to remove the negative values. One additional consideration is that the count of missing tokens must be taken into account while computing $c$ and $p$. The following equations consolidate the ideas: 
%\todo{I recommend we use the backslash notation for the multiset difference.}
\begin{equation}
    c \gets \text{sum}(\mathbf{M} \setminus  \mathbf{M'}) + m
\end{equation}
\begin{equation}
p \gets \text{sum}(\mathbf{M'} \setminus \mathbf{M}) - m
\end{equation}

%\begin{equation}
%    c \gets \text{sum}({\text{max}{\left((\mathbf{M} -\mathbf{M'}), \mathbf{O}\right)})} + m
%\end{equation} 
%\begin{equation}
%p \gets \text{sum}({\text{max}{\left((\mathbf{M'} - \mathbf{M}, \mathbf{O}\right)})} - m
%\end{equation}

The number of remaining tokens $r$ is computed only when the net reaches the final marking. When that happens, $r$ is $|M' \setminus f|$ with $f$ being the final marking. Since $f \subseteq M'$, because we require the final marking to be reached, the result using vector difference will not contain any negative value. However, this will only be true when we reach the final marking. To ensure that this has happened, we use an approach similar to the Equation~\ref{eq:selector}. First, let us compute min($\mathbf{M}', \mathbf{1}$), which is equivalent to turning the multiset into a set (the multiset may have values greater than 1, so we replace such values by a 1). Then, we compute de the product $\mathbf{f}^T \cdot\text{min}(\mathbf{M}^\prime, \mathbf{1})$ which will result into an scalar. Such scalar is the number of coincidences in both vectors. If all the numbers that correspond to the places in the final marking are found in the marking $\mathbf{M}'$, then the count of such places will be equal to sum($\mathbf{f}$), which we use as a divisor. Hence, the expression sum($\mathbf{f}^T\cdot \text{min}((\mathbf{M}', \mathbf{1}) \div \text{sum}(\mathbf{f}))$) will be 1 if all places in the final marking have at least a token. Otherwise, the same expression becomes 0. With that in mind, we can now define the following:
\begin{equation}
    r \gets \text(sum)(\mathbf{M}' - \mathbf{f})\cdot\text{sum}(\mathbf{f}^T\cdot \text{min}((\mathbf{M}', \mathbf{1}) \div \text{sum}(\mathbf{f})))
\end{equation}

\subsection{The algorithm}

The equations described in the previous subsections capture the steps required to implement token-based replay. These steps are integrated in Algorithm~\ref{alg:step}. Recall the method would be executed using a client/server architecture. Lines 1-16 corresponds to code for the server side and the rest, i.e. lines 17-29, are those for the client side.

\begin{algorithm}
\setstretch{1.25}
\begin{algorithmic}[1]
\Statex --------------- Server side ------------
\Function{Init}{$\mathcal{N}$}
\State {\bf global} $\ \mathcal{E},\ \mathcal{S},\ \mathcal{P} \gets$ \Call{ComputeMatrices}{$\mathcal{N}$}
\State {\bf global} $\ \mathbf{f} \gets$ \Call{ComputeFinalMarking}{$\mathcal{N}$}
\EndFunction

\Function{CheckFinalMarking}{$\mathbf{M}', \mathbf{M}$}
% ----- REMAINING
\State $f \gets \text{sum}((\mathbf{f}^T \cdot \text{min}\mathbf{(M',1)}) \div \text{sum}(\mathbf{f}))$
\State $r \gets \text{sum}\mathbf{(M' - f)} \cdot f$
\State \Return $f, r$
\EndFunction

\Function{Step}{$\mathbf{M}$, $\mathbf{t}$}
\LeftComment \emph{Is there a firing sequence $\boldsymbol{\sigma} =\tau^* t$ enabled by marking $\mathbf{M}$?}
\State $\mathbf{\hat{M}} \gets \text{min}\mathbf{(M, 1)}$
%\left[\begin{matrix} \text{min}(\mathbf{M}[i,0], 1) \end{matrix}\right] : 0 \leq i < |\mathbf{M}|$
\State $\mathbf{sel} \gets\ \left(\boldsymbol{\mathcal{E}}\cdot \left[\begin{matrix}
    \mathbf{\hat{M}}\\
    \mathbf{t}^T
\end{matrix}\right]\right) \div \boldsymbol{\Delta}$
\State $\boldsymbol{\sigma} \gets\ \sum_{i = 0}^ {|\mathbf{sel}|-1} \boldsymbol{\mathcal{S}}[i] \cdot \mathbf{sel}[i]$
\LeftComment \emph{Compute the preset of transition $t$}
\State $\boldsymbol{\pi} \gets\ \sum_{i = 0}^{|\mathbf{t}|-1} \boldsymbol{\mathcal{P}}[i] \cdot \mathbf{t}[i]$
% \State Preset $\gets\ \sum_{0 \leq i < |\mathbf{t}|} \text{Presets}[i] \cdot \mathbf{t}[i]$
% \Statex
% \State Selection $\gets \sum_{i = 0}^{|\text{Selector}|}\text{Selector}[i]$
\State $\mathbf{M'} \gets \left(\mathbf{M} + \mathbf{N} \cdot \boldsymbol{\sigma}^T \right)\cdot \text{sum}(\mathbf{sel}) + \left(\mathbf{M} + \boldsymbol{\pi}^T + \mathbf{N} \cdot \mathbf{t}^T\right)\cdot \left(1-\text{sum}(\mathbf{sel})\right)$

% \Statex
% \LeftComment \emph{Compute local counters $c,m,p,r$ to compute fitness}
% ----- MISSING
\State $m \gets \text{sum}(\boldsymbol{\pi}) \cdot (1 - \text{sum}(\mathbf{sel}))$

\State \Return $\mathbf{M'},m$
\EndFunction
\Statex
\Statex --------------- Client side ------------
\Function{ReplayTrace}{$\theta$}
\State $\mathbf{M} \gets \mathbf{0}$ \Comment{All zeroes marking}
\State $P,C,Mi,R \gets 0,0,0,0$
\For {$e \in \theta$} \Comment{For each event in the trace}
\State $\mathbf{t} \gets$ \Call{Encode}{e}
\State {$\mathbf{M'},m \gets$ $\mathbb{DEC}$(\Call{Step}{$\mathbb{ENC}(\mathbf{M, t})$})}
\State $P \gets P + \text{sum}(\mathbf{M'} \setminus  \mathbf{M}) - m$
\State $C \gets C + \text{sum}(\mathbf{M} \setminus  \mathbf{M'}) + m$

\State $Mi \gets Mi + m$
\State $\mathbf{M_{prev}},\mathbf{M} \gets \mathbf{M, M}'$
\EndFor
\State $f, r \gets \Call{CheckFinalMarking}{\mathbf{M, M_{prev}}}$
\State $R \gets f = 1\ ?\ r : \text{sum}(\mathbf{M'})$
\State \Return $\frac{1}{2}\left(1 - \frac{Mi}{C}\right)
        + \frac{1}{2}\left(1 - \frac{R}{P}\right)$
\EndFunction
\end{algorithmic}
\caption{Secure Token-based Replay}
\label{alg:step}
\end{algorithm}

Let us illustrate how the replay of a trace occurs using Algorithm~\ref{alg:step}. First, the server side is initialized by providing the net to be considered during conformance checking, by calling the function {\sc Init} in lines 1-4. Such function initializes all the matrices and the final marking, and stores them in global variables.

On the client side, the trace owner calls the function {\sc ReplayTrace} with a trace $\theta$. The marking vector is initialized with all elements as zeroes, line 18. The global token counters are also initialized to zero, line 19. Then, it enters a loop for processing each event in the trace at a time. In line 21, the vector $\mathbf{t}$ is computed by calling a function {\sc Encode}. That function is expected to produce a vector with all zeroes except for the element that represents the transition. It is assumed that the client side and the server side use the same positions in the vectors to represent the transitions. The name of the event must have a counterpart transition in the net. In line 22, the function {\sc Step} is called with the marking and the vector represents the transition to be executed. However, before calling the function {\sc Step} we have to encrypt the vector $\mathbf{M}$ and $\mathbf{t}$ by calling the function {$\mathbb{ENC}$}\footnote{Client and server side must exchange a public keys in order to enable homomorphic encryption. We omit such technicalities and focus only on the algorithmic aspects.}. Now it is the turn of the server side to execute the function {\sc Step}. It easy to see that the body of {\sc Step} comprises all the equations that we explained in the previous subsections.%, so we consider that the reader can already understand what happens there. 
At the end, {\sc Step} returns the new marking and the local values for counters $p,c,m,$ and $r$. The computation in {\sc Step} is performed with homomorphic operations. Moreover, the values returned by {\sc Step} are encrypted. Hence, the client side proceeds by decrypting the results using the function $\mathbb{DEC}$. In lines 23-26, the traces values for $P,C,Mi,\text{ and } R$ are updated. When all the events in the trace have been processed, we compute the trace fitness in line 28 and return it as the result.

The temporal complexity of Algorithm~\ref{alg:step} is dominated by the matrix multiplications in lines 7, 8, 9, and 10. Among them, $\mathcal{E}$ is the larger matrix. Let $\rho$ be the number of enablements of the net, which variates depending on the topology of the Petri net. On the other hand, $\mathcal{E}$ has $\kappa = |P|+|T|$ columns, that is the sum of the number of places plus the number of transitions. Hence, the matrix multiplication in line 7 requires $O(\rho \cdot \kappa)$ homomorphic operations.
However, in practice the time required to perform homomorphic operations over encrypted data remains the most critical factor when it comes to estimate the execution times. Particularly, the precision of the encrypted values. For instance, the matrix multiplication in line 12 can be done over unsigned (small) integers. In a first version of Algorithm~\ref{alg:step}, our {\sc Step} function included the computation of counters $p,c,m$ and $r$. However, that decision resulted in very long execution times, i.e., more than 30 minutes, which we attribute to the use of integers of 16 bits of precision that, once encrypted, induces very expensive computations.

Since homomorphic encryption is said to be not deterministic: a given value (a.k.a., plaintext) can be associated with different encrypted values (a.k.a., ciphertext). The latter makes it difficult for the model's owner to get to know the details about the trace being processed. On the other hand, the model's owner can use a scheme (e.g., circular shifting the elements in the vector) to obfuscate the marking, to hinder the trace's owner to infer the structure the model (beyond its own trace). Such obfuscation is, however, not considered in Algorithm~\ref{alg:step}.

\section{Evaluation}
\label{sec:evaluation}

We implemented our algorithm for secure token-based replay using Zama´s Concrete framework~\cite{Concrete}. The code and datasets used for our evaluation are publicly available~\footnote{\url{https://github.com/lgbanuelos/SToRe}}. As mentioned before, Concrete allows one to specify homomorphic computation in plain Python, i.e., a Python function, which is later compiled and linked to a native library. Concrete wraps up some of the matrix/vector operations from NumPy, making our implementation very concise and even similar to the pseudocode in Algorithm~\ref{alg:step}.

For the evaluation, we implemented four different versions of the prototype. First, we implemented the part that performs the token-based replay and omitted the computation of the token counts: one version working with clear values, referred to as CLR, and another version working with encrypted data, referred to as SEC. Further, we implemented two additional versions that computes also the token counts. We refer to them as CLR+ and SEC+, respectively.

With these prototypes, we evaluated the efficiency of the method. The experiments were run on a MacBook Air M1 2020, equiped with 16GB of RAM, and running the operating system MacOS Sonoma 14.6.1.

For our evaluation, we used the model and the event log of our running example, which are taken from the book~\cite{Aalst16} and the tutorials of PM4Py\footnote{\url{https://github.com/process-intelligence-solutions/pm4py/blob/release/notebooks/4_conformance_checking.ipynb}}. The Petri net was preprocessed so as to produce all the matrices and vectors beforehand, which we stored in a JSON file. Such JSON file was later loaded from our Python code. The time taken for this preprocessing step is on the order of seconds.

We took a set of traces from the event log consisting of 4 traces with 5 events, one trace with 9 events and another one with 13 events. As expected, CLR and CLR+ terminated in milliseconds. Token-based replay without token counts was completed by SEC in 7.8 to 15.8 secs. Finally, the full version SEC+ required between 19.3 and 37.4 seconds to complete. Hence, in all the cases the conformance checking over encrypted information took less than 1 minute. We repeated the experiments with another 6 traces, included in a log with ``broken'' traces, because 5 out of the 6 were unfitting, with lengths of 4 to 12 events. The times were consistent with those of fitting traces, with a maximum of 23 seconds for the longest trace.

We implemented a version where the token counters $p, c, m$ and $r$ are computed in the function {\sc Step}. Surprisingly, the execution times of this version were extremely long with values between 35 to 84 minutes. To our understanding, the differences in the times are associated with the overhead induced by the size of the values exchanged as input and as output to the function that uses homomorphic encryption. Concrete requires one to provide random samples to determine the ranges of the values in the vectors provided as input to the function Step. Hence, we can provide an upper bound to the values in the marking. We used a bound of 6, hence, Concrete tunes the compilation to use 3-bits integers. However, Concrete does not provide a similar approach to tune the size of the values to be returned. Our conjecture is that the compiler defaults the counters $p,c,m$ and $r$ to 16-bits integers which, when encrypted, result in very large representations. All the homomorphic operations for computing and returning the values of $p,c,m$ and $r$ have to be adjusted to such underlying representation. 

Overall, the execution times of the secure token-based replay are considerably larger than the times working on clear data. However, we consider them still as acceptable in contexts were secure computation is a must.

% To evaluate our proposed algorithm Step~\ref{alg:step} four different versions were implemented.

% \begin{itemize}
%     \item Clear-mcpr. Direct implementations of algorithm Step~\ref{alg:step}
%     \item Clear. Implementations of algorithm Step~\ref{alg:step} without the computation of the missing, consumed, produced and remaining tokens.
%     \item Cipher-mcpr. Encrypted implementations of algorithm Step~\ref{alg:step}
%     \item Cipher. Encrypted implementations of algorithm Step~\ref{alg:step} without the computation of the missing, consumed, produced and remaining tokens.
% \end{itemize}

% The token replay algorithm was used to verify the conformance of the Petri net of the figure~\ref{fig:running-example}

% \begin{figure}
%     \begin{subfigure}[b]{1in}
%         \includegraphics[width=6cm]{figs/tr_enc.pdf}        
%         \label{fig:times_chart_enc}
%         %\caption{Execution times for the encrypted token replay}
%     \end{subfigure}\hfil
%     \begin{subfigure}[b]{1in}
%         \includegraphics[width=6cm]{figs/tr_clear.pdf}        
%         \label{fig:times_chart_clear}
%         %\caption{Execution times for the plain text token replay}
%     \end{subfigure}    
%     \label{fig:times_chart_enc}
%     \caption{Execution times for the encrypted and plain text token replay version proposed in this paper.}
    
% \end{figure}

\section{Conclusion and future work}
\label{sec:conclusion}

Conformance checking is one of the main operations in process mining. In this operation, discrepancies between the expected and observed process behaviour are computed by comparing a process model (expected behaviour) against an event log (observed). There are different approaches to conformance checking, one of them is token-based replay. This approach is inspired by the token game, which is often used for discussing the execution semantics of a Petri net. In this approach, discrepancies between the process model and the event log are given by means of missing and remaining tokens.

Traditionally, the process model and the event log are both accessible to the business analyst performing the conformance checking. However, in some contexts it is necessary not to disclose the model or to keep the log secret to protect critical or sensitive information. This paper presents, for the first time, a secure conformance checking using homomorphic encryption. The evaluation, performed using a synthetic event log, shows the practicality of the proposed technique in scenarios where secure computation is a must. 

In future work, we will evaluate this approach using real-life event logs. Another direction of future work is to address the token flooding problem in token-based replay. Additionally, our current prototype works in a single block and it is intended as a proof of concept. We want then to re-implement it, possibly using other platforms to further analyze the trade-offs between precision of the number representations and performance. In addition, we plan to implement a client-server version of the algorithm to also analyze the costs of communication.
%The implementation in these frameworks requires an analysis of the impact on performance considering the cost of communication in client-server settings.

\subsection*{Acknowledgments}
This work has received funding from the Swiss National Science Foundation under Grant No. IZSTZ0 208497 (ProAmbitIon project). Astrid Rivera-Partida is doctoral fellow with CVU 1007189, funded by Mexico’s government (CONAHCyT).

\bibliographystyle{splncs04}
\bibliography{refs}
\end{document}